\documentstyle[twocolumn,aps,epsfig]{revtex}

\newcommand{\be}{\begin{eqnarray}}
\newcommand{\ee}{\end{eqnarray}}

\begin{document}
\draft

\title{
The Origin of Large-$p_T$ $\pi^0$ Suppression at RHIC}
\author{Sangyong Jeon$^{1,2}$, Jamal Jalilian-Marian$^3$ and Ina Sarcevic$^4$}
\address{
$^1$RIKEN-BNL Research Center, Upton, NY 11973-5000\\
$^2$Department of Physics, McGill  University, Montreal, 
QC H3A-2T8, Canada \\
$^3$Physics Department, Brookhaven National Laboratory, Upton, NY 11973, USA 
 \\
$^4$Department of Physics, University of Arizona, Tucson, Arizona
85721, USA\\
}

\wideabs{

\maketitle

\begin{abstract}

\widetext 

We present results for inclusive $\pi^0$ production in proton-proton and in 
Au-Au at RHIC energy $\sqrt s=200$ GeV.  We use next-to-leading order 
perturbative 
 QCD calculation and we include 
 nuclear effects such as parton energy loss and 
nuclear shadowing. We consider the ratio of $\pi^0$ distribution in Au-Au 
and p-p collisions for $p_T > 3$ GeV  for three cases of parton energy loss: 
1) constant parton energy loss per parton scattering, $\epsilon^a_n=const$, 
2)  Landau-Pomeranchuk-Migdal energy-dependent energy loss, 
$\epsilon^a_n \sim \sqrt E^a_n$ and 3) Bethe-Heitler energy-dependent 
energy loss, $\epsilon^a_n \sim E^a_n$.  We show that recently observed 
suppression of $\pi^0$ production in Au-Au collisions at RHIC, which is 
found to increase  with $p_T$ increasing from $3$GeV to $8$GeV, can be 
reproduced by $\epsilon^a_n=0.06 E^a_n$. We show that the ratio of prompt 
photons to neutral pions produced in Au-Au collisions at RHIC has a strong 
$p_T$ dependence approaching one at $p_T\sim 10$GeV.  

\end{abstract}
}

\vskip 0.1true in

\narrowtext

Recent results from the Relativistic Heavy-Ion Collider (RHIC) 
showing large suppression of $\pi^0$ production in Au-Au 
collisions relative to proton-proton collision, in the 
large $p_T$ region ($3$GeV $<p_T<8$GeV), have 
created new excitement in the field \cite{phenix}.  
Since partons produced in hard parton-parton collisions propagate 
through the hot and dense medium created in the
heavy-ion collision and lose their energy, production of large 
$p_T$ pions could potentially provide a unique opportunity for 
studying the properties of the hot and dense matter and the possible 
formation of a new phase, the quark-gluon plasma \cite{wang1,hdsg,jjs}. 
Furthermore, parton energy loss is expected to suppress the spectrum 
of the final state pions produced in heavy-ion collision as compared 
to pions produced in hadronic collisions \cite{wang}. The observed 
$p_T$ dependence of the suppression was quite unexpected \cite{phenix}  
and posed a great challenge for theoretical models which typically 
predicted a rise of this ratio with $p_t$ at large transverse
momentum \cite{wang}. 

The aim of this letter is to show that the observed suppression can be 
reproduced in an economical way using next-to-leading order pQCD 
calculation of inclusive $\pi^0$ production \cite{se} augmented 
with energy dependent energy loss and nuclear shadowing. We emphasize 
that NLO pQCD \cite{se} predictions for $\pi^0$ production in 
proton-proton collisions at RHIC energy of $\sqrt s=200$ GeV was found 
to be in good agreement with PHENIX data \cite{phenix} for $p_T > 3$ GeV
without a need for introduction of parton intrinsic transverse momentum.
We show that assuming that $6\%$ of parton energy is lost per each parton 
scattering, one can reproduce the observed $p_t$ dependence of the 
suppression. Here we also present results for prompt photon production in 
proton-proton and in Au-Au collisions at RHIC energy of $\sqrt s=200$ GeV
and discuss the $p_T$ region in which prompt photons become dominant.  

In perturbative QCD, the inclusive cross section
for pion production in a hadronic collision is given by: 

\be
E_\pi \frac{d^3\sigma}{d^3p_\pi}(\sqrt s,p_\pi)
&=&
\int dx_{a}\int dx_{b} \int dz \sum_{i,j}F_{i}(x_{a},Q^{2}) \nonumber \\
&&
F_{j}(x_{b},Q^{2}) D_{c/\pi}(z,Q^2_f) E_c 
{d^3\hat{\sigma}_{ij\rightarrow c X}\over d^3p_c} 
\label{eq:factcs}
\ee

\noindent
where $F_{i}(x,Q^{2})$ is the i-th parton distribution in a nucleon,
$x_a$ and $x_b$ are the fractional momenta of incoming partons,
$D_{c/\pi}(z,Q^2_f)$ is the pion fragmentation function, $z$ is the fraction 
of parton energy carried by the pion and 
${d^3\hat{\sigma}_{ij\rightarrow c X}\over d^3p_c}$ are parton-parton cross 
sections which include leading-order, $O(\alpha_s^2)$, subprocesses 
and the next-to-leading order, $O(\alpha_s^3)$, subprocesses.  

\begin{figure}[htp]
\centering
\setlength{\epsfxsize=8.5cm}
\centerline{\epsffile{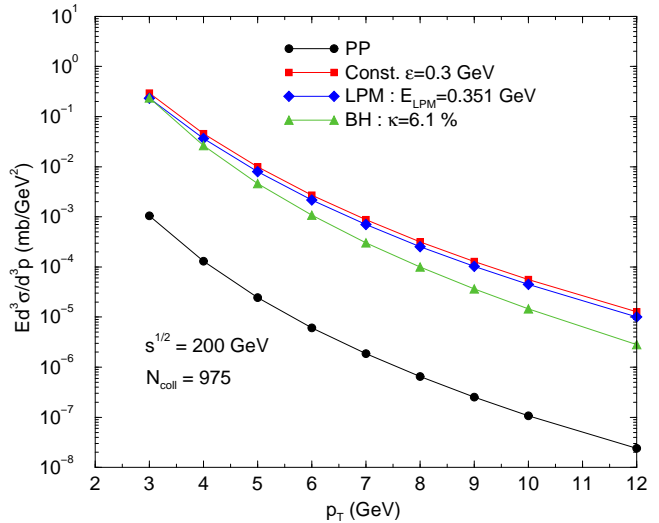}}
\caption{Inclusive $\pi^0$ cross section in proton-proton and in 
Au-Au 
 collisions at $\sqrt s=200$ GeV.}
\label{fig:rhic200dsigma_new}
\end{figure}

The parton distribution functions $F_{i}(x,Q^{2})$ are measured in Deep
Inelastic Scattering experiments at HERA \cite{hera}, while
fragmentation functions, $D_{c/\pi}(z,Q^2_f)$, that describe the transition 
of the partons into the final-state pions are extracted from $e^+e^-$ 
annihilation data from PETRA, PEP and LEP \cite{frag}.  
We use MRS99 
 parameterization of nucleon structure functions 
\cite{mrs99}, which includes NLO corrections, 
BKK pion fragmentation functions \cite{frag}, 
and we set renormalization, factorization and fragmentation scale to 
be equal to $p_T$.  Varying scale from $p_T$ to $2p_T$ does not 
affect the shape of the distribution, but rather the overall normalization 
 \cite{se,jji}.  
Our results for inclusive $\pi^0$ production in proton-proton collision 
at RHIC energy is shown in Fig. \ref{fig:rhic200dsigma_new}. Recent 
PHENIX data on $\pi^0$ production in proton-proton collision at 
$\sqrt s=200$GeV is found to be in agreement with NLO pQCD predictions 
\cite{phenix}. Since we restrict ourselves to $p_t > 3$ GeV region, we
do not include the so called parton intrinsic momenta which are
expected to be rather small in the high $p_t$ region \cite{papp}.  
                                                          
To calculate the inclusive cross section for pion production in heavy
ion collisions, we will use (\ref{eq:factcs}) with the distribution
and fragmentation functions appropriately modified to include nuclear
effects such as shadowing and energy loss.
The modification of the parton distribution, 
known as nuclear shadowing (for a review see 
 \cite{arneodo}), 
 can be written as 
\begin{eqnarray*}
F_{a/A}(x,Q^2,b_t)=T_A(b_t)\,S_{a/A}(x,Q^2)\,F_{a/N}(x,Q^2)
\end{eqnarray*}
\noindent
where $T_A(b_t)$ is the nuclear thickness function, $F_{a/N}(x,Q^2)$ is the
parton distribution function in a nucleon and $S_{a/A}(x,Q^2)$ is the
parton shadowing function.  
We use recent parameterization of the shadowing function of 
 Eskola, Kolhinen and Salgado (EKS98), 
which is $Q^2$ dependent, distinguishes between quarks and gluons 
\cite{eks98} and is shown to be in very good agreement with the NMC data 
on $Q^2$ dependence of $F_2^{Sn}/F_2^C$ \cite{NMC}.  

We also include the medium induced parton 
energy loss effect.  Fast partons produced in parton-parton collision 
propagate 
 through
the hot and dense medium and through scatterings 
 lose
part of their energy \cite{eloss} and then fragment into hadrons with 
a reduced energy. 
While a dynamical study of the parton propagation in a hot and dense medium
created in a realistic heavy-ion collision and the modification of the 
hadronization is more desirable, we will use a simple phenomenological model 
\cite{hsw} here to demonstrate sensitivity of pion production to the 
parton energy loss.  
 Given the 
inelastic scattering mean-free-path, $\lambda_a$, the probability for a 
parton type $a$ to scatter $n$ times within a distance $\Delta L$ before it escapes 
the system is assumed to be given by the Poisson distribution.  
The modified 
fragmentation function is given by \cite{hsw}, 

\be
% \sum_{n=0}^NP_a(n) z^a_n D^0_{a/\pi}(z^a_n,Q^2)
z D_{a/\pi}(z,\Delta L,Q^2)& =&
 \sum_{n=0}^N P_a(n)
 \Bigg[
 z_n^a D^0_{c/\pi}(z_n^a, Q^2)  
  \nonumber \\
%&+&\langle n_a\rangle z'_aD^0_{g/\pi}(z'_a,Q_0^2), 
&& +
 \sum_{m=1}^n z_m^a D_{g/\pi}^0(z_m^a, Q^2) % the sum should be inside
 \Bigg]
\label{eq:mfrag}
\ee

\noindent
where 
$z_n^a = z E^a_T/E^a_n$, 
$E_{n+1}^a = E_n^a - \epsilon^a_n$, 
$z_m^a = p_T/\epsilon^a_m$ and 
 $D^0_{a/\pi}$ is the hadronic 
fragmentation function which gives the probability that quark or a gluon 
would fragment into a pion. 
The average number of scatterings within a distance $\Delta L$
is $\langle n_a\rangle =  \Delta L/\lambda_a$.  We take 
$\lambda_a=1fm$ and $\Delta L= R_A$.  

We calculate the invariant cross section for $\pi^0$ production in heavy-ion 
collision normalized 
to the number of binary nucleon-nucleon collisions, $N_{coll}$, 
where $N_{coll}$ can theoretically be 
determined from nuclear 
overlapping function, i.e.  
$N_{coll}= T_{AA}(b)\sigma_{inel}^{NN}$.  
The number of N-N collisions 
depends on the centrality that experiment triggers on.  Here we take 
$N_{coll}=975$, which is obtained 
by PHENIX for their central collisions \cite{phenix}.  
In order to investigate the sensitivity of our results to the choice of
energy loss parameters in the model of \cite{hsw}, we consider three cases 
of parton energy loss.  
 First 
we take energy loss to be constant in each parton scattering, 
$\epsilon^a_n=const.$, then we consider 
LPM energy-dependent energy loss \cite{eloss}, 
$\epsilon^a_n = \alpha_s\sqrt{E^a_n E_{LPM}}$ and BH energy-dependent energy loss, 
$\epsilon^a_n = \kappa E^a_n$.  
The scale separating the LPM region of validity from the BH region is 
 given by the product of the 
average $<p_T^2>$, $p_T$ being the kick that parton gets from each collision and 
 the mean free path, $\lambda_a$.  Taking $<p_T^2>$ to be $1 GeV^2$ and $\lambda_a$ to
 be 1 fm, one gets the scale to be about 5 GeV.
 Above this scale, one would expect LPM effect to be dominant.  
However, increasing the value of $\lambda_a$ or $<p_T^2>$ 
 shifts the transition place 
to higher values of $p_T$.  Furthermore, one should keep in mind that 
LPM effect in QCD has been derived for static scatterers \cite{eloss}, which 
may not be suitable approximation in case on hot QGP. We find that the
parton energy loss is mostly responsible for the observed suppression
of hadron spectra in heavy ion collisions and that nuclear shadowing 
is a small effect ($<10\%$ at most). Furthermore, the $p_T$ spectra are 
quite insensitive to the choice of scales \cite{jji}.  
We show our results for the inclusive $\pi^0$ production in Au-Au 
collisions at $\sqrt s=200$ GeV in Fig. \ref{fig:rhic200dsigma_new} 
for different choice of parton energy loss parameter, $\epsilon^a_n$.  
We chose the value for the constant energy loss, $\epsilon^a_n=0.3$GeV,  
$E_{LPM}=0.35$GeV, and the fraction of the energy lost per scattering, 
$\kappa = 6\%$, such that the suppression at $p_T = 3$ GeV is 
consistent with the observed suppression for $p_T\le 3 $GeV \cite{rhic}.  
Then we predict the shape of $p_T$ distribution for $p_T>3$ GeV for different 
choices of energy loss.  We note that $p_T$ distribution of neutral pions 
is much steeper in case of BH energy-dependent parton energy loss.  

\begin{figure}[htp]
\centering
\setlength{\epsfxsize=8.5cm}
\centerline{\epsffile{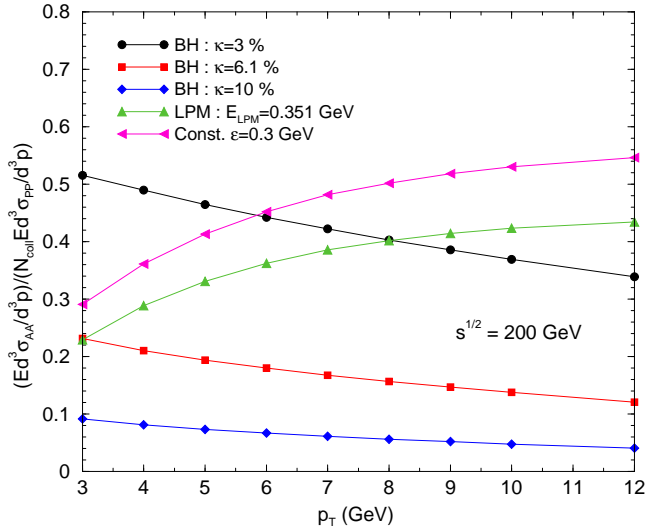}}
\caption{Ratio of inclusive $\pi^0$ cross sections in Au-Au 
collision and in proton-proton collision at $\sqrt s=200$ GeV.}
\label{fig:rhic200ratio_new}
\end{figure}

In Fig. \ref{fig:rhic200ratio_new} we show the ratio of $\pi^0$ 
production in Au-Au collisions to the one in p-p collisions for 
different choice of parton energy loss parameter, $\epsilon^a_n$.  
We find that for constant energy loss and for LPM energy-dependent 
energy loss, the ratio increases with $p_T$, while for the BH case, 
the ratio slightly decreases with $p_T$.  We show that the ratio is 
very sensitive to the fraction of energy loss, for example, for 
$\kappa=10\%$, it decreases from about $50\%$ to $35\%$ for $p_T$ 
increasing from $3$ GeV to $12$GeV, while for $\kappa =3\%$ it 
decreases from $90\%$ to $96\%$.  When $\kappa=6\%$, the suppression 
increases from $76\%$ at $p_T=3$ GeV to $84\%$ at $p_T=8$GeV, in 
agreement with recent PHENIX data \cite{rob}.  
Clearly measurements of inclusive pion production at RHIC energies for 
$p_T>3$ GeV have provided valuable information about the medium induced 
parton energy loss since nuclear shadowing effects are very small 
(few $\%$) and the observed suppression of $\pi^0$ in Au-Au collisions 
can be described with an energy-dependent parton energy loss. However,
theoretical expectations that a LPM type of energy loss should be 
present is clearly not seen for $p_T < 5-6$ GeV while LPM type energy
loss seem to work for higher $p_T$.  One should keep in mind that
all theoretical calculations of LPM energy loss are for very energetic 
partons and at high $p_T$ and applicability to RHIC at low to 
intermediate $p_T$'s is not appropriate.  BH type of energy loss 
seem to reproduce the data quite well up to $p_T \sim 8-10$ GeV 
 but will 
fail beyond that. The most interesting aspect of this is that there 
seems to be different energy loss mechanisms at work at different $p_T$'s. 
Understanding $\pi^0$ production at RHIC is also important because 
large-$p_T$ $\pi^0$ pions form a significant background for the prompt 
photons. 

\begin{figure}[htp]
\centering
\setlength{\epsfxsize=8.5cm}
\centerline{\epsffile{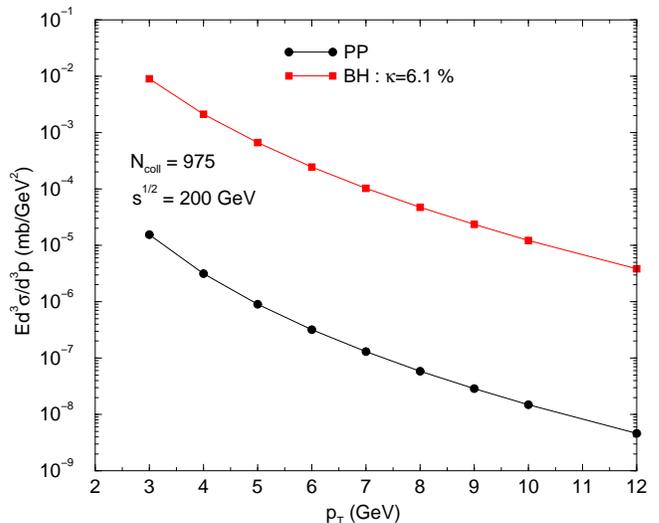}}
\caption{Prompt photon production in proton-proton and in Au-Au collisions 
at $\sqrt s=200$ GeV.}
\label{fig:rhic200dsigma_ph_new}
\end{figure}

\begin{figure}[htp]
\centering
\setlength{\epsfxsize=8.0cm}
\centerline{\epsffile{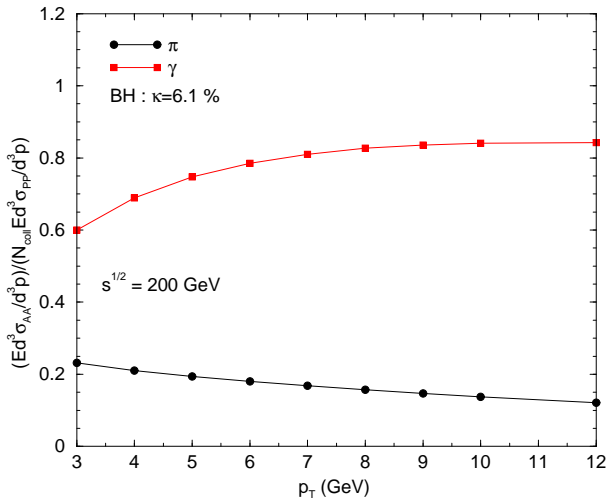}}
\caption{Ratio of prompt photons produced in Au-Au collisions and in 
proton-proton collision at 
$\sqrt s=200$ GeV, compared with the same ratio for neutral pions.}
\label{fig:rhic200dgratio_pratio_new}
\end{figure}

\begin{figure}[htp]
\centering
\setlength{\epsfxsize=8.0cm}
\centerline{\epsffile{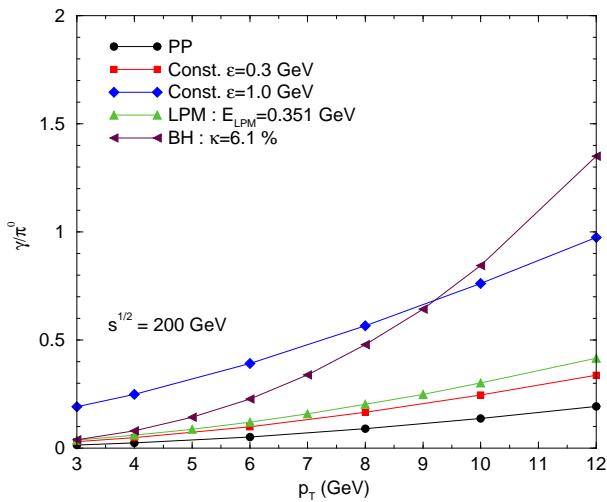}}
\caption{Ratio of prompt photon and $\pi^0$ cross sections at 
$\sqrt s=200$ GeV.}
\label{fig:rhic200_goverp_new}
\end{figure}

Studying photon production at the RHIC is also of special interest, 
as it has been suggested as an elegant signal for detecting the formation 
of a quark-gluon plasma (QGP) in heavy-ion collisions \cite{mt}. Photons 
can be produced at different stages of the heavy-ion collision:  
photons produced at the early stages of the collision through 
hard parton-parton scatterings or they can be emitted from a thermalized 
quark gluon plasma or hadronic gas. Furthermore, prompt photons are an 
important background to thermal photons in the low to intermediate
$p_t$ region and are dominant in the high $p_t$ region. Therefore, it is
extremely important to be able to calculate their production cross section
reliably. Here we present results for prompt photon production in 
proton-proton and in Au-Au collisions at RHIC energy of $200$GeV 
using the NLO pQCD code of Aurenche et al.  with nuclear shadowing and 
medium induced parton energy loss effects included. We use the same 
initial parton distributions, choice of scales, nuclear shadowing 
function and choice of parton energy loss, $\epsilon^a_n=0.06E^a_n$ as 
for pions.  However, we note that in prompt photon production there are 
two types of subprocesses that contribute, direct processes and 
bremsstrahlung processes, the later one being convoluted with the 
fragmentation function \cite{se}. Fragmentation functions,
$D_{c/\gamma}(z,Q^2_f)$, that describe the transition of the partons into
the final-state $\gamma$, without medium effects, are extracted from 
$e^+e^-$ data \cite{fragg}. In Fig. \ref{fig:rhic200dsigma_ph_new} we show 
our results for prompt photon distribution in proton-proton and in 
Au-Au collisions at $\sqrt s=200$GeV.  We note that suppression of 
prompt photons produced in heavy-ion collisions is much less that for 
$\pi^0$.  This is due to the fact that only bremsstrahlung processes 
are affected by the parton energy loss, which contribute $24\%$ to 
the cross section at $p_T=3$ GeV and $6\%$ at $p_T=12$ GeV. For the 
same reason, prompt photons are not very sensitive to a different 
choice of parton energy loss.  
In Fig. \ref{fig:rhic200dgratio_pratio_new} we show the ratio of prompt 
photon production in Au-Au and in proton-proton collision compared to 
the same ratio in case of $\pi^0$ production. We take $\epsilon^a_n$ to 
be energy-dependent, $\epsilon^a_n=0.06E^a_n$.  We find that the ratio 
for prompt photon production is not very sensitive to the choice of energy 
loss, and for all choices considered, constant parton energy loss and 
energy-dependent parton energy loss, the suppression decreases with 
increasing $p_T$. We find that while the $\pi^0$ suppression increases 
from $76\%$ to $88\%$ for $p_T$ increasing from $3$ GeV to $12$ GeV, 
prompt photon suppression decreases from $40\%$ to $15\%$ in the same 
range of $p_T$.  
In Fig. \ref{fig:rhic200_goverp_new}, we show the ratio of prompt photons 
to $\pi^0$ as a function of transverse momentum for $\sqrt s=200$ GeV, 
including nuclear shadowing and energy-dependent parton energy loss, 
$\epsilon^a_n = 0.06 E^a_n$. We also show this ratio for proton-proton 
collisions at the same energy.  In Au-Au collisions at RHIC, because of the 
large $\pi^0$ suppression relative to prompt photons at RHIC energies, this 
ratio increases with $p_T$ approaching $1$ at $p_T \sim 10$ GeV. 
In summary we have investigated inclusive pion and prompt photon production
in proton-proton and in Au-Au collisions at RHIC. We have used the NLO
pQCD results of Aurenche et al. for pion and prompt photon production in
proton-proton collisions and included nuclear shadowing and medium induced 
parton energy loss by modifying the final state pion and photon fragmentation 
functions.  

We find  nuclear shadowing effects to be rather small while medium 
induced parton energy loss results in large suppression of $\pi^0$ 
production in Au-Au collisions at $\sqrt s=200$ GeV. We show that 
the suppression is sensitive to the type of energy loss assumed.  
In the case of constant or LPM energy loss, the suppression decreases 
with increasing $p_T$.  On the other hand, when partons traversing the 
medium lose constant fraction of their energy, the suppression becomes 
stronger at higher $p_T$, which seems to be in agreement with recent 
PHENIX data \cite{phenix}. The fact that one can reproduce the data
with our simple and quite naive model of energy loss is remarkable. 
Indeed, there are now more sophisticated models which include 
 effects such as intrinsic transverse momentum of incoming partons, 
$p_T$ broadening due to multiple parton scatterings as well as effects 
due to finite energy of the partons. These models also lead to the 
suppression of pion production in Au-Au collisions, however, the 
suppression is either 
decreasing with $p_T$ \cite{wangwang} or is almost $p_T$ independent for 
$2 GeV < p_T <20 GeV$ at RHIC energies \cite{vitev}. The effects of 
dynamical expansion of the collision region have been recently studied 
and related via scaling law to an equivalent static scenario 
\cite{wiedemann}.   The advantage of our simplistic model is that 
 it can reproduce the data with only a 
few model-dependent assumptions.  

Using our results for pion production, we predict the prompt photon 
production cross section in proton-proton and Au-Au collisions at RHIC, 
where we use an energy-dependent parton energy loss, 
$\epsilon^a_n=0.06E^a_n$ and EKS nuclear shadowing function. We also 
present the ratio of prompt photon and $\pi^0$ production 
at RHIC, of relevance to separating different sources of photon production.  
At RHIC energy we find that at $p_T=3$GeV direct processes contribute 
about $75\%$, while bremsstrahlung processes are remaining $25\%$.  
This changes significantly as one goes to higher energies, such as LHC, 
where bremsstrahlung contribution becomes dominant and photons become 
as suppressed as pions \cite{jji}.  

\leftline{\bf Acknowledgments}
We are indebted to P. Aurenche and J. P. Guillet for providing us with
the fortran routines for calculating 
$\pi^0$ and photon 
production in hadronic collisions.  
We thank D. d'Enterria, M. Tannenbaum and 
S. Mioduszewski 
for many helpful 
 suggestions. 
I.S.  is supported in 
part through U.S. Department of Energy Grants Nos. DE-FG03-93ER40792 and 
DE-FG02-95ER40906. S.J. is supported in part by the Natural Sciences and
 Engineering Research Council of Canada.  
J.J-M. is supported in part by a PDF from BSA and by 
U.S. Department of Energy under Contract No. DE-AC02-98CH10886.

\end{document}